\documentclass[11pt,twocolumn]{revtex4}   % style for Physical Review B and AJP are similar

\usepackage{amsmath}    % need for subequations
\usepackage{graphicx}   % for figures
\def\ba{\begin{eqnarray}}
\def\ea{\end{eqnarray}}
\def\be{\begin{equation}}
\def\ee{\end{equation}}

\begin{document}

\title[The Locus of the apices]{The Locus of the apices of projectile trajectories under constant drag}
%Lines break automatically or can be forced with \\
\author{H. Hern\'andez-Salda\~na}
%\email{hhs@correo.azc.uam.mx}   %optional
% \altaffiliation[Also at ]{home.}  %  optional
\address{Departamento de Ciencias B\'asicas,\\
Universidad Aut\'{o}noma Metropolitana at Azcapotzalco,\\
Av. San Pablo 180, M\'{e}xico 02200 D.F., Mexico.}
%\author{Jan Tobochnik}
%\affiliation{Kalamazoo College, Department of Physics, Kalamazoo,
%MI 49007}
\date{\today}

\begin{abstract}
We present an analytical solution for the projectile coplanar motion under constant drag parametrised 
by the velocity angle. We found the locus formed by the 
apices of the projectile trajectories.
The range and time of 
flight are obtained numerically and we find that the optimal launching angle is smaller than in the free drag 
case. This is a good example of problems with constant dissipation of energy that includes curvature, 
and it is proper for intermediate courses of mechanics. 
\end{abstract}

\maketitle
%%%%%
%\begin{multicols}{2}
\section{Introduction} 

Projectile trajectory under constant  drag has deserved a lot of attention in
literature, not only as it appears as a common problem in undergraduate physics 
and can be given recent exact analytical results and new analysis\cite{Yabushita2007,Belgacem2014, Mustafa2016a,Morales2016,Stewart2012}. 
The power-law velocity dependent drag 
\be
\vec{f} = -\sum_n mg b_n v^n \dfrac{\vec{v}}{v},
\label{pldrag}
\ee
with $v=||\vec{v}||$, is a series approximation for the real complex problem. The linear, $n=1$, and quadratic, $n=2$, cases are of much used, not only for the analysis of the motion of  a particle in midair but 
as well to model other energy dissipation process. In the quantum scales, $n=1$ is a usual model for the 
energy losses\cite{Dittrich1998quantum,razavy2006classical}.

Notwithstanding the usefulness of linear approximation, and that allows analytical solutions for the projectile motion, equation (\ref{pldrag}) have another case: $n=0$, the constant drag case.  It is not trivial if, as it is usual, the projectile motion is coplanar, i.e., the vector $\vec{v}/v$ changes with the orientation of the orbit.\footnote{In one dimension this term only changes the sign of the force in order to keep it in opposition to the movement.} 
This case has deserved few attention, the only reported work we found are \cite{Jones1991,Jones1991a}.

There is not evidence that exists a regime where the drag could be considered constant, however
the problem studied here is important for the following reasons: 

i)  A series expansion for a retarding force has to have a no null zeroth term, take for instance,  the 
integrable Legendre cases
\be
f (v) = \dfrac{1}{n}(a+b v^n),
\ee
where the constant $a \neq 0$ appears\cite{MacMillan}. 
 ii) The motion of an object in a non-newtonian fluid with yield stress could be constant, see for
 instance \cite{Barnes1999}, i.e., the problem of a particle launched in oil or liquid chocolate contains this constant term. Even more, spheres into loose granular media are another example of an object moving in a  fluid with presence of yield stress\cite{Bruyn2004}.  
 iii) As an undergraduate problem, a  constant retarding force could be considered as a point 
 rocket with the thrust pointed against the motion. iv) As a simple example of friction that depends on 
 the curvature.      

In the present paper we analyse such a case, obtaining both the explicit solutions of the problem in the next section and  the
description of the locus which give title to this work. We discuss the range and the time of flight are given in section
\ref{rangetime}. Conclusions are presented in section \ref{conclusions}.

\begin{figure}[tb]
\includegraphics[width=0.8 \columnwidth]{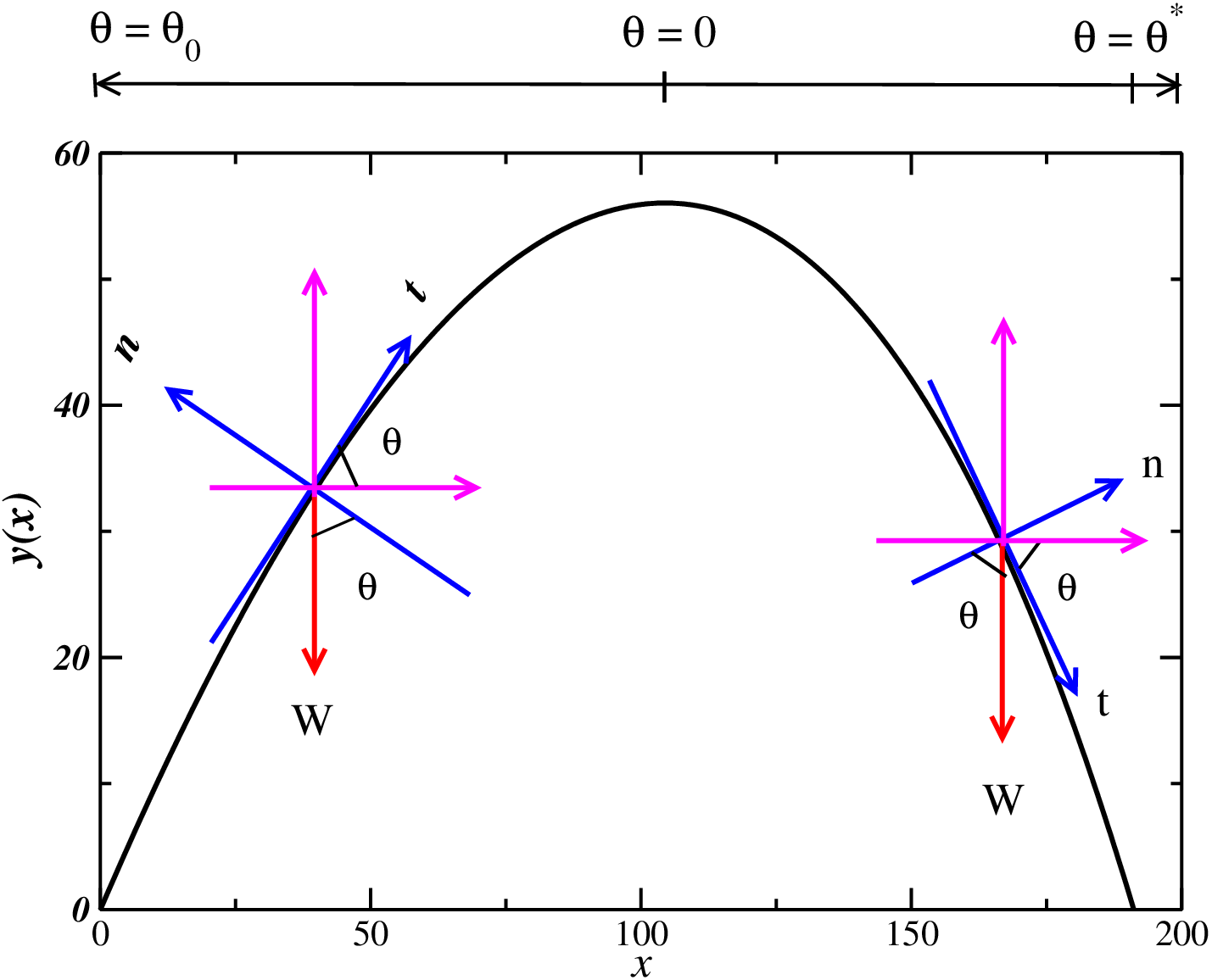}
\caption{(color online) The normal, $n$, and tangent, $t$, reference frame during the projectile movement (in blue lines) as comparison to the usual reference frame $x-y$ (in magenta lines). The first comparison previous to reaching the apex  and the second, afterward. In the upper scale we show the velocity angle scale $\theta$ starting at $\theta_0$ up to the angle when $y=0$, i.e, $-\theta = \theta ^* $. The final, asymptotic, angle is $\theta = - \pi/2$, regardless the value of the dissipation parameter. As usual, $W = mg$, indicated in red. Plotted trajectory corresponds to $v_0 = 50$, $\theta_0 = \pi/4$ and $b_0 = 0.25$ in SI units.} 
\label{Fig:frame}
\end{figure}
     
\section{The projectile problem with constant drag}\label{Projectilemotion}

The constant drag problem is governed by the following equations in rectangular coordinates

\begin{eqnarray}
m \frac{d^2}{dt^2} \vec{r} = -mg \hat{\jmath} -mgb_0 \vec{v}/v.
\end{eqnarray}
Notice that the friction is constant {\it in the direction of motion}, i.e., it changes with velocity. We choose the drag force in units of weight $mg$ in order to compare with linear and
quadratic drag results. 

In order to be clear on what kind of differential equations we are dealing with, we explicitly rewrite 
the above equations in cartesian coordinates 
\ba
\begin{array}{lll}
m \ddot{x} & = & -mgb_0 \dfrac{\dot{x}}{\sqrt{\dot{x}^2+\dot{y}^2}}, \\
m \ddot{y} & = & -mg -mg b_0 \dfrac{\dot{y}}{\sqrt{\dot{x}^2+\dot{y}^2}}. 
\label{cdrag}
\end{array}
\ea
Here, we use a dot for a time derivative. The above equations are coupled and non-linear. However an analytical solution parametrised with  the velocity angle can be obtained. Some
 other results require of standard numerical methods\cite{Burden}. The solutions presented here for $x$ and $y$ do not requiere of any further numerical integration.
%   Implementation of \texttt{Dsolve} in \texttt{Mathematica}\cite{Mathematica} offers an explicit long solution (see Fig. XXX) with the same problems of inversion as we shall see with the hodograph solution presented in the next section. In order to draw the solutions as a function of time a standard four degree Runge-Kutta numerical integration is used. 

\section{Explicit solution parametrised by $\theta$.}

In order to obtain a solution of the problem (\ref{cdrag}) we first change the equations for  normal, $n$, and tangent, $t$,   coordinates to the  motion, hence, the corresponding force components are
\begin{equation}\label{tang-force}
F_t = -mg \sin \theta -mg b_0, 
\end{equation}
and
\begin{equation}\label{norm-force}
F_n = -mg \cos \theta.
\end {equation}
If the mass is constant, we obtain 
\begin{equation}\label{tang-acel}
m \dot{v} = -mg \sin \theta -mg - mg b_0
\end{equation}
and
\begin{equation}\label{norm-acel}
m \frac{v^2}{\rho} = -mg \cos \theta.
\end {equation}
where $\rho = -d s/ d \theta$ and, $s$ is the arc length. The last equation 
can be written as 
\begin{equation}\label{vdetheta}
v \dot{\theta} = -g \cos \theta,
\end{equation}
with the help of the chain rule: $\rho = -d s/ d \theta = - (d s/d t) (d t/d \theta )$. Equation (\ref{tang-acel}) for the tangent acceleration can be modified with the same rule and using (\ref{vdetheta}) the result is 
\begin{equation}\label{master0}
\frac{d v}{d \theta} = v (\tan \theta + b_0 \sec \theta).
\end{equation}
For the initial conditions $v(t=0) = v_0 $ and $\theta(t=0) = \theta_0$,
we solve this first order differential equation obtaining 
\begin{equation}\label{vtheta}
v(\theta) = \frac{v_0 \cos \theta_0 }{\cos \theta } \left(\frac{\Delta}{ \Delta_0} \right),
\end{equation}
with  
$$\Delta \equiv (\sec \theta + \tan \theta)^{b_0}, $$
and $\Delta_0 \equiv \Delta(\theta_0)$.
%$$\Delta_0 \equiv (\sec \theta_0 + \tan \theta_0)^{b_0}.$$.

The solution for time is
\begin{eqnarray}\label{textheta}
\begin{array}{lll}
t(\theta) & = &  -\dfrac{1}{g} \int_{\theta_0}^{\theta} v(\theta) \sec \theta d \theta \\
      &  = &   -\dfrac{v_0 \cos \theta_0 }{g \Delta_0 }
        \Big(  \dfrac{(b_0- \sin \theta) \Delta}{(b_0^2-1) \eta}    
         -   \dfrac{(b_0- \sin \theta_0) \Delta_0}{(b_0^2-1)\eta_0} \Big),
\end{array}   
\end{eqnarray}
being 
\be
\eta = (\cos \theta/2 -\sin  \theta/2 ) (\cos \theta/2 + \sin  \theta/2 ),
\ee
and $\eta_0 \equiv \eta(\theta_0)$.

%The hodograph solution

Using a similar procedure we obtain 
%Since $\dot{x} = v \cos \theta$ and $\dot{y} = v \sin \theta$, we can obtain the solutions via the 
%following integrals
\begin{widetext}
\begin{equation}
\begin{array}{lcl}
x(\theta)  &  = & -\frac{1}{g} \int_{\theta_0}^{\theta} v(\theta)^2 d \theta \\
    &   = & -\dfrac{1}{ g }\left(\dfrac{v_0 \cos \theta_0} {\Delta_0} \right)^2  %\\
%  &   &
  \Big[ -\dfrac{(-2 b_0 + \sin \theta) \Delta^2}{(2 b_0-1)(2b_0+1) \eta}+  % \\
%  &   &    \\
 % &   &    
  \dfrac{(-2 b_0 + \sin \theta_0) \Delta_0^2}{(2 b_0-1)(2 b_0+1)\eta_0} \Big],
\end{array}  
\label{xexthe}
\end{equation}
\end{widetext}
and
\begin{widetext}
\begin{eqnarray}
\begin{array}{lcl}
y(\theta)  &  = & -\frac{1}{g} \int_{\theta_0}^{\theta} v(\theta)^2 \tan \theta d \theta \\
        & = & -\dfrac{1}{ g }\left(\dfrac{v_0 \cos \theta_0} {\Delta_0}\right) ^2  %\\
%  &   &
  \Big[\dfrac{\sec^2 \theta \left(-3+\cos 2 \theta +4 b_0 \sin\theta \right) \Delta^2}{8 (b_0^2-1)} \\
  &  & -\dfrac{\sec^2 \theta_0 \left(-3+\cos 2 \theta_0 +4 b_0 \sin\theta_0 \right) \Delta_0^2}{8 (b_0^2-1)}\Big].
\end{array}
\label{yexthe}  
\end{eqnarray}
\end{widetext}
So,  (\ref{textheta}),(\ref{xexthe}) and (\ref{yexthe}) are, formally, the solutions to  the problem (\ref{cdrag}). Unfortunately, 
explicit inversion of $t(\theta)$ is hard (if not imposible). 
Notwithstanding, these solutions are analytical and no additional integration is requiered.
In search of an explicit time dependent solution homotopy analysis method could offer a 
guide as it was the case of quadratic drag \cite{Yabushita2007}. 

\begin{figure}[tb]
\includegraphics[width=\columnwidth]{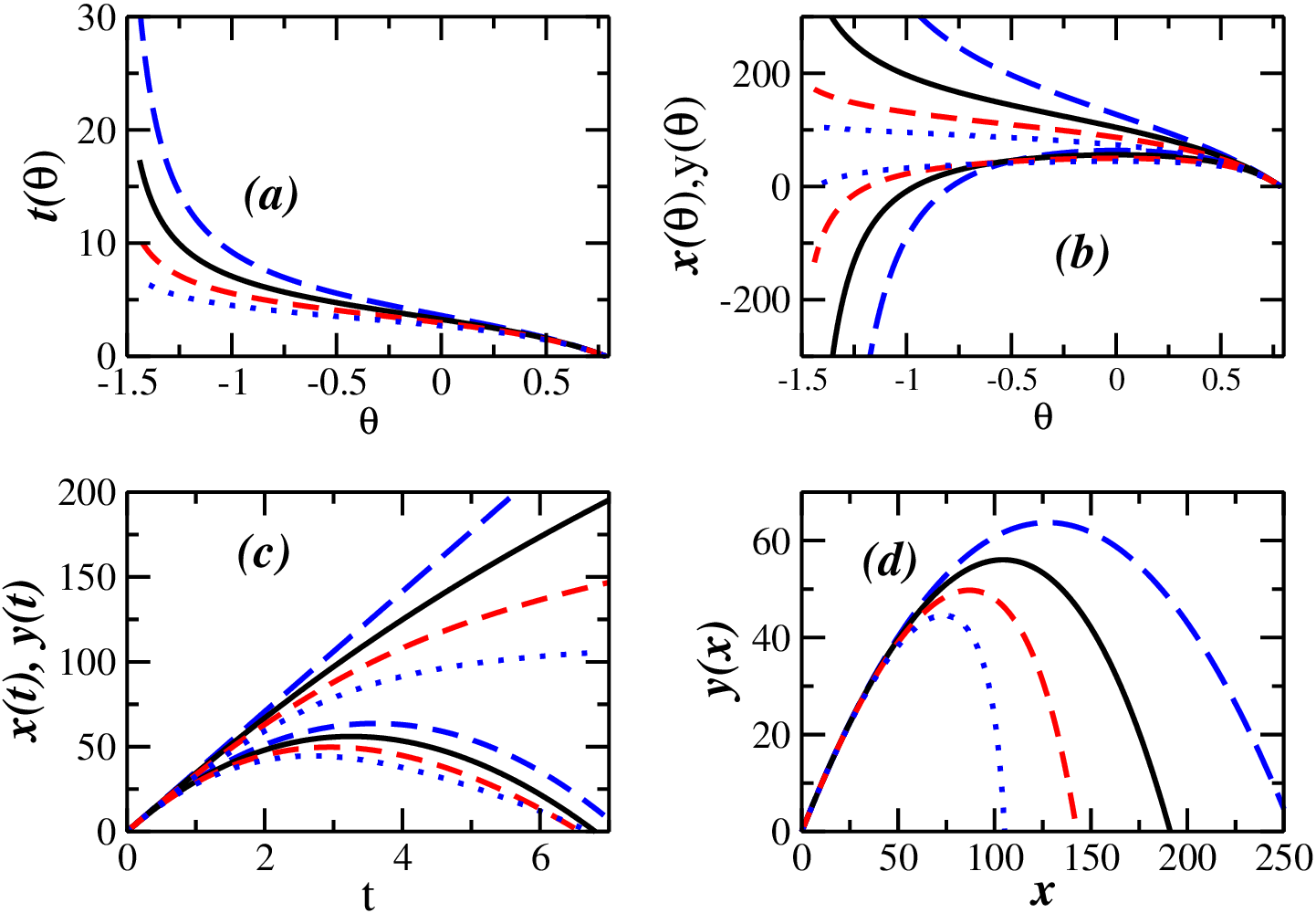}
\caption{(color online) Explicit solution to the projectile motion in presence of a constant drag as function of the angle $\theta$, in all cases we draw the drag-free solution in blue broken lines, $b_0 = 0.25$ in black lines, $b_0 = 0.5$ in red dashed lines and the blue dotted lines correspond to $b_0=0.75$.  In (a) time is depicted, notice that the angle go from $\theta_0$ to $-\pi/2$ as time increases. In (b) we shown $x$ and $y$ as function of the angle, functions that growing to $\infty$ correspond to $x$ and those going to $-\infty$ are for $y$ solutions. In (c) and (d) we depict the solution in the  traditional variables, $x$ and $y$ as function of time for the former and $y$ as function of $x$ in the latter.}
\label{Fig:sol}
\end{figure}

In order to establish that previous expressions are as useful as the time parametrisation we shall use them to plot the 
usual graph of $x(t)$ and $y(t)$ as well as  the iconic  $y(x)$(see \ref{Fig:sol}). 
For comparison, we rewrite the free drag solutions as function of $\theta$, the results 
\begin{equation}
\label{tfree}
t(\theta) = -\frac{v_0 \cos \theta_0}{g} \left( \tan \theta - \tan \theta_0 \right),
\end{equation}
\begin{equation}
\label{xfree}
x(\theta) = -\frac{(v_0 \cos \theta_0)^2}{g} \left(\tan \theta - \tan \theta_0 \right),
\end{equation}
and
\begin{equation}
\label{yfree}
y(\theta) = -\frac{(v_0 \cos \theta_0)^2}{2 g} \left(\sec^2 \theta - \sec^2 \theta_0 \right),
\end{equation}
are obtained by 
solving (\ref{tang-acel}) and (\ref{norm-acel}) for $b_0 = 0$.
It is an exercise to check that the previous  expression are the familiar solutions of parabolic motion.

First we explain the solutions in angle parametrisation. To this end 
we draw equation  (\ref{textheta}) in figure \ref{Fig:sol}(a), i.e.,  time as function of $\theta$ for $b_0 = 0.25$ (black line), $b_0 = 0.5$ (red dashed line) and $b_0 = 0.75$ (blue dotted line) and the free drag case in blue dashed line, from (\ref{tfree}). The launching angle was set to $\theta_0 = \pi/4$ here, other selection shall
shift the graphs (not shown). The parameter $\theta$ go, asimptotically, to $-\pi/2$, since the reference frame change the orientation after the orbit reach its apex as it appears in figure \ref{Fig:frame}.

In figure \ref{Fig:sol}b) solutions (\ref{xexthe}) for $x(\theta)$ are presented in the same order as before (graphs diverging to  $\infty$ as $\theta \to -\pi/2$) .  The solutions (\ref{xexthe}) for  $y(\theta)$ are those that diverge to $-\infty$ as $\theta \to -\pi/2$. A close up of them (not shown) could show the angle where 
$y=0$. The numerical solutions to this condition shall be discussed below. Again we draw in blue-dashed lines the drag free solutions from (\ref{xfree}) and (\ref{yfree}).

%from, in figure (\ref{Fig:sol})(b) $x(\theta)$ and $y(\theta)$ from (\ref{xexthe}) for $b_0 = 0.25$ (black line), $b_0 = 0.5$ (red dashed line) and $b_0 = 0.75$ (blue dotted line).  

% meanwhile $y \to -\infty$ in the same limit. In blue-dashed lines the drag free solutions are depicted in all frames and they correspond to solutions (\ref{tfree}), (\ref{xfree}) and (\ref{yfree}).  

In figure (\ref{Fig:sol})c) time solutions are presented for  $x(t)$ (upper graphs) and $y(t)$ (lower graphs).
Using (\ref{xexthe}), (\ref{yexthe})  and a simple computational program we can write the 
 $x(t(\theta))$ and  $x(t(\theta))$ data and plot it.  We do that and we present the results for the same drag values and color code. We consider only the range of $\theta$ in order to show the 
 $y=0$ condition.   The $y$ results show the larger the drag the shorter the maxima. The maxima are reached at a shorter times as the drag increases, as well. 
 
 Finally, we present the iconic $y(x)$ for projectile motion in  figure (\ref{Fig:sol})d). As expected, the 
 larger the $b_0$ value the shorter the path. Certainly, at first sight the paths are similar to those obtained 
 with a linear drag, but a comparison require to compare energy losses, not similar values of 
 $b_0$ and $b_1$\cite{hhsx}.
      
% the $y(x)$ appears in figure (\ref{Fig:sol})d) with 
%the characteristic lost of range for increasing value of $b_0$. 

% \begin{equation}
%\begin{array}{ccc}
%t(\theta) & = &-\frac{v_0 \cos \theta_0 }{g (\sec \theta_0 + \tan \theta_0)^a}
 %       \big(  \frac{(a- \sin \theta) (\sec \theta + \tan \theta)^a}{(a-1)(a+1)(\cos \theta/2 -\sin  \theta/2 ) (\cos \theta/2 \sin  \theta/2 )} \\
%  & & \\
 %       &  & - \frac{(a- \sin \theta_0) (\sec \theta_0 + \tan \theta_0)^a}{(a-1)(a+1)(\cos \theta_0/2 -\sin  \theta_0/2 ) (\cos \theta_0/2 \sin  \theta_0/2 )} \big)
%\end {array} 
%\end{equation}

\subsection{The locus of the apices}\label{locusapices}

\begin{figure}[tb]
\includegraphics[width=0.9 \columnwidth]{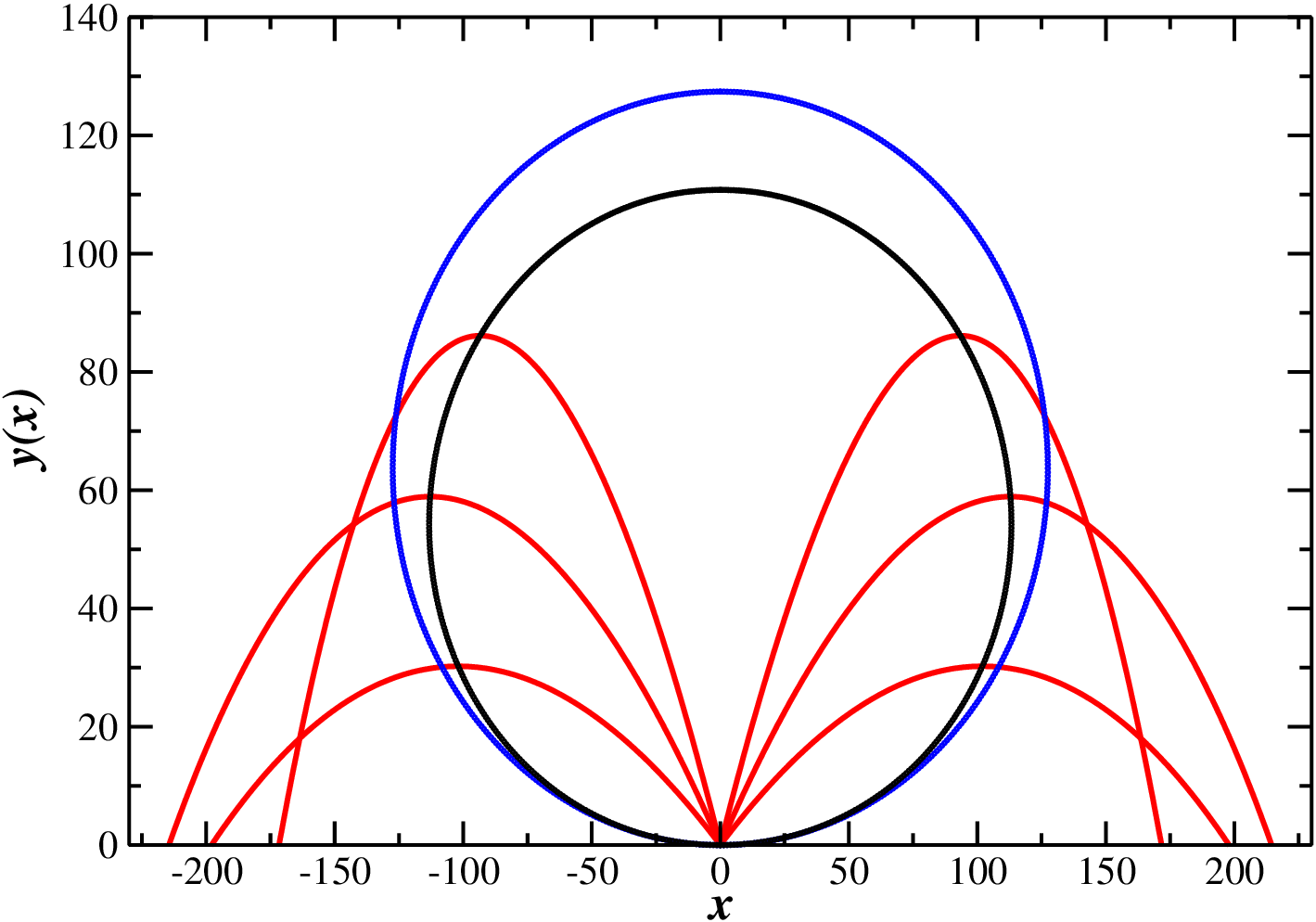}
\caption{(color online) Locus formed by the apices of all the projectile trajectories for constant drag with the same initial speed (black line). In blue broken lines the solution for the free drag case. The parameters used here are
$v_0 = 50$, $b_0=0.15$ and we plot in red the trajectories for $\theta_0 = 30^{\circ}$,  $\theta_0 = 45^{\circ}$ and
$\theta_0 = 60^{\circ}$.} 
\label{Fig:locus}
\end{figure}

The solution in terms of the angle could be hard to handle but gives a straightforward 
for a particular locus: the locus formed by all the apices for initial launching angle $\theta_0$.
The cases for no drag\cite{Salas2004,ThomasCalculus,Salas2014} and linear drag has been studied previously\cite{Stewart2006, Stewart2011,hhs2010}.

The apex for each orbit is obtained by setting $\theta = 0$ for $x$ and $y$ in (\ref{xexthe}) and (\ref{yexthe}) as can be seen at figure (\ref{Fig:frame}).  After rearranging factors in these equations and using $(\cos \theta_0/2 -\sin  \theta_0/2 ) (\cos \theta_0/2 + \sin  \theta_0/2 )= \cos \theta_0$\footnote{ Since the launching angle remain in the first quadrant}, the 
locus is written  as 
\begin{widetext}
\begin{eqnarray}
x(\theta_0) & = &-\frac{1}{g(4 {b_0}^2-1)} \left(\frac{v_0 \cos \theta_0 }{\Delta_0} \right)^2 \Big[
2 b_0  +  \frac{(-2 b_0 + \sin \theta_0) {\Delta_0}^{2}}{\cos \theta_0 } \Big]
\end{eqnarray}
\end{widetext}
and 
%\begin{eqnarray}
\begin{widetext}
\be
\begin{array}{lll}
y(\theta_0)  &  = & \dfrac{1}{8 g ({b_0}^2-1)} \left(d\frac{v_0 \cos \theta_0} { \Delta_0} \right)^2   \Big[ 2 %\\
%  &  &
   + \sec^2 \theta_0 \left(-3+\cos 2 \theta_0 +4 b_0 \sin\theta_0 \right) {\Delta_0}^2 \Big].
\end{array}
\ee
\end{widetext}
%\end{eqnarray}

%%%%%
%\begin{eqnarray}
%x(\theta_0) & = &\frac{1}{g} \left(\frac{v_0 \cos \theta_0 }{(\sec \theta_0 + \tan \theta_0)^{b_0}} \right)^2 
%\Big[
%\frac{(2 b_0 )}{(2b_0-1)(2b_0+1)}   \\
%  &   &  +  \frac{(-2 b_0 + \sin \theta_0) (\sec \theta_0 + \tan \theta_0)^{2 b_0}}{(2 b_0-1)(2 b_0+1)(\cos \theta_0/2 -\sin  \theta_0/2 ) (\cos \theta_0/2 + \sin  \theta_0/2 )} \Big]
%\end{eqnarray}
%and 
%\begin{eqnarray}
%y(\theta_0)  &  = & \frac{1}{ g }\left(\frac{v_0 \cos \theta_0} {(\sec \theta_0 + \tan \theta_0)^{b_0}} \right)^2   \Big[\frac{ 1}{4 (b_0^2-1)} \\
%  &  & - \frac{\sec^2 \theta_0 \left(-3+\cos 2 \theta_0 +4 b_0 \sin\theta_0 \right) \left((\sec \theta_0 + \tan \theta_0)^{b_0} \right)^2}{8 ({b_0}^2-1)}\Big]
%\end{eqnarray}
%%%%%%%
In figure (\ref{Fig:locus}) we show the locus for parameters with values $v_0 = 50$ m/s, $b_0= 0.15$ and $g = 9.81$ m/s$^2$. 
The drag-free solution
\be
x_m = \rho \sin \theta_0 \cos \theta_0,
\ee
\be
y_m = \frac{\rho}{2}\sin^2 \theta_0,
\ee
is shown for comparison. We add three orbits, those corresponding to launching angles $\theta_0 = 30^{\circ}$,  $\theta_0 = 45^{\circ}$ and
$\theta_0 = 60^{\circ}$ as is usual in the textbooks.

% as well as the linear case,
%\be
%x_m = \frac{\rho \cos \theta_0 \sin \theta_0}{1 + \epsilon \sin \theta_0},
%\ee
%\be
%y_m = \frac{\rho}{\epsilon^2}\left( \epsilon \sin \theta_0 - \log (1 + \epsilon \sin \theta_0)\right),
%\ee
%for comparison.  Here $\rho =v_0^2/g$ and $\epsilon =b v_0/g $ for linear drag force as $-m b \vec{v}$.

%Notice that the linear drag produces shorter orbits for this values, however if we set the initial speed in order to obtain similar drag forces, i.e. 
%$-m b v_0$ and $-mga$ the locus are similar (see inset in figure XXX).

\subsection{Some important quantities in projectile motion: The range and the flight time}\label{rangetime}

\begin{figure}[tb]
\includegraphics[width=0.9 \columnwidth]{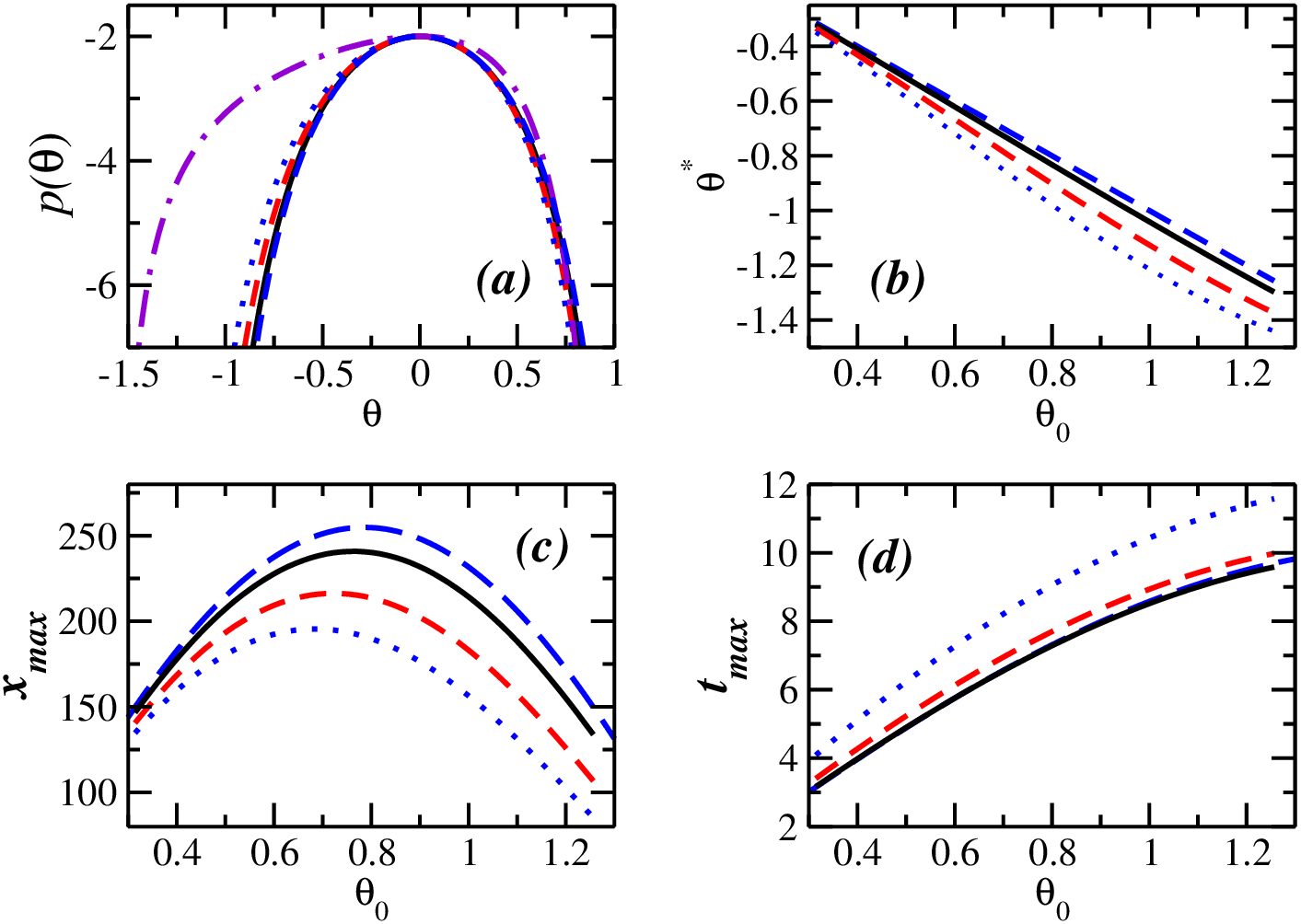}
\caption{(color online) Range, $x_{max}$, and flying time  } 
\label{Fig:range}
\end{figure}

Unfortunately not all the important quantities are of mathematical significance. Meanwhile the apex has a mathematical meaning, other locus are important for practical reason. Such is the case of the 
range and its maximum. 
Their value are determined by our choose of the origin and the chord generated.  The selection of the origin is determined in an arbitrary way and hence the length of the chord. 
Hence, it is not surprising that we need to 
solve numerically (\ref{yexthe}) for $y = 0$. 

This condition is translated from (\ref{yexthe}) to
solve the equation
\begin{widetext}
\begin{equation}
\label{fth}
\sec^2 \theta \left(-3+\cos 2 \theta +4 b_0 \sin\theta \right) \Delta^2
 = \sec^2 \theta_0 \left(-3+\cos 2 \theta_0 +4 b_0 \sin\theta_0 \right) \Delta_0^2
\end{equation}
\end{widetext}
If we call  $p(\theta)= \sec^2 \theta \left(-3+\cos 2 \theta +4 b_0 \sin\theta \right) \Delta^2$, we are looking for solutions such that $p(\theta) = p(\theta_0)$. For 
symmetrical functions the solutions is clear, but this is not the case  as can be seen in figure \ref{Fig:range}a). 
There we plot $p$ for the indicated values of $b_0$ and the free-drag case $\sec^2 \theta (-4 + 2 \cos^2 \theta )=  sec^2 \theta_0 (-4 + 2 \cos^2 \theta_0 )$ with the solution $\theta^* = \pm \theta_0$ (in blue dashed line). 
In this figure the drag values considered are $b_0=0.05$ in black line, $0.15$ in red dashed line  and $0.25$ in dotted blue line. We add the extreme case of $b_0 = 0.75$ in order to show how asymmetric the curve $p(\theta)$ can be\footnote{ Calculations for  values  of 
$b_0$ larger than $0.25$ require of a better selection of the initial condition as we can be seen for $
b_0 = 0.75$ in figure \ref{Fig:range}(a)}. The color code remains in the rest of the graphs.

In figure  \ref{Fig:range}b)
we show the solution obtained via  Newton-Raphson for the equation $p(\theta ^*)-p(\theta_0)=0$  and the corresponding case for the range $x_{max}=x(\theta ^*)$ as function of the launching angle in figure \ref{Fig:range}c).

%In figure \ref{Fig:range}(a), $p(\theta)$ is plotted  for $b_0=0.05, 0.15$ and $0.25$. Two functions 
%were added for comparison: The drag free solution from (\ref{yfree}) where we added a $-2$ factor to 
%keep the same limit when $b_0 \to 0$ and an extreme case of $b_0 = 0.75$. In \ref{Fig:range}(b) the results
%for the angle $\theta^*$ as function of launching angle $\theta_0$. The drag free solution is $\theta^* = -\theta_0$.  

In the last figure the maximum range 
occurred at $\theta_0 \approx 0.7697$,$0.7226$ and $0.6912$ for the  indicated values of $b_0$.  All these values are smaller than $\theta_0 = \pi/4 \approx 0.7854$, the corresponding value for the drag free case (shown in blue broken line). 
For completeness, we present the time of flight as function of the launching angle  in figure \ref{Fig:range}(d). Such a time 
increases with the angle. Notice that the drag free case and the solution for $b_0 = 0.05$ are so close that they appear superimposed.

\section{Conclusions}\label{conclusions}

We discussed the motion of a projectile under the influence of constant gravitational pull and constant
drag. Such a case could be considered as the yield  stress in a non-newtonian fluid and as an example of
a simple situation where the retarding force depends on the velocity direction. 
%Since the motion change in direction the drag change in accordance, giving 
The two coupled non-linear differential equations in rectangular coordinates can be exactly solved by 
a change to normal and tangent coordinates. The solutions, (\ref{xexthe}) and (\ref{yexthe}), are parametrised as functions of the velocity angle. That allow us to obtain the locus of the apices in an explicit way. Other locus or quantities requiere of numerical calculation as the range and flight time presented in the previous section.
 
%A change of reference frame to the tangent and normal to the trajectory allows a 
%closed solution in term of a the angle between the the velocity and the horizontal axis instead of time. 
%Even when inversion of equation (\ref{textheta}) is not evident, the formal solution is explicitly  written 
%in (\ref{xexthe}) and (\ref{yexthe}). That allow us to obtain the locus of the apices in an explicit way.
%Other locus or quantities requiere of numerical calculation.

This problem serves as a good example for introduce undergraduate students to problems with curvature and retarding forces, 
beyond the problem of an inclined plane with constant friction.
\section*{Acknowlegment}
We thank to AL Salas-Brito for valuable comments. 

\section*{References}
\bibliography{projectilemotion}

%\end{multicols}{2}
\end{document}